\documentclass[prl,twocolumn,showpacs,amsmath,amssymb]{revtex4}
\usepackage{dcolumn}
\usepackage{bm}
\usepackage[dvipdf]{graphicx}
\usepackage[dvips]{color}
\DeclareGraphicsExtensions{.jpg,.pdf,.mps,.png,.eps,.ps,.EPS}

\begin{document}

\def\be{\begin{equation}}
\def\ee{\end{equation}}
\def\bc{\begin{center}} 
\def\ec{\end{center}}
\def\bea{\begin{eqnarray}}
\def\eea{\end{eqnarray}}
\newcommand{\avg}[1]{\langle{#1}\rangle}
\newcommand{\Avg}[1]{\left\langle{#1}\right\rangle}

\title{Local structure of directed networks}

\author{Ginestra Bianconi$^1$, Natali Gulbahce$^2$, and Adilson E. Motter$^3$}
\affiliation{$^1$The Abdus Salam International Center for Theoretical 
Physics, Strada Costiera 11, 34014 Trieste, Italy\\
$^2$Theoretical Division and CNLS, Los Alamos 
National Laboratory, NM 87545, USA\\
$^3$Department of Physics and Astronomy and NICO,
Northwestern University, IL 60208, USA}

\begin{abstract}
Previous work on undirected small-world networks established the paradigm that
locally structured networks tend to have high density of short loops. On the
other hand, many realistic networks are {\it directed}. Here we investigate the
local organization of directed networks and find, surprisingly, that real
networks often have very 
few short loops as compared to random models. We
develop a theory and derive conditions for determining if a given network has
more or less loops than its randomized counterparts.  These findings carry broad
implications for structural and dynamical processes sustained by directed networks.
\end{abstract}

\maketitle 

Asymmetric interactions are widespread in natural and technological networks,
particularly when the network transports a flow or underlies collective behavior \cite{Klemm}. The 
structure of such directed networks can be characterized
by the statistics of loops, the building blocks of closed paths, which provides 
information on structural correlations \cite{Alex}, motifs, robustness and redundancy
of pathways, and impacts dynamical as well as equilibrium critical phenomena on the
network \cite{Dorogovtsev}.   

In undirected networks, the large number of short loops
together with small diameter gives rise to the small-world
effect encountered in many real systems \cite{swn}. Strikingly, 
in this Letter we show
that there is a large class of directed networks for which the number of loops
is strongly reduced with respect to the random hypothesis. The directed 
neural network of {\it C.\ elegans}, for example, has less than 50\% of the short loops expected
from a random ensemble with the same degree sequence despite the well-known fact that, when regarded as an
undirected network \cite{swn}, it has a clustering coefficient 5.6 times larger
than randomly rewired versions of the network.

Motivated by this empirical finding, we demonstrate numerically and analytically that
degree correlations \cite{corr} strongly constrain the loop structure of directed networks.
Moreover, we go beyond the degree-correlated picture and derive conditions for determining
if a {\it given} network has more or less loops than its randomized counterparts. We characterize
the network local organization in terms of short loops and its global organization
in terms of long loops. We compare our analytical results with exact (when possible)
or approximate numerical calculation of the number of loops  in a class of directed networks
that includes foodweb, power-grid, metabolic, neural, transcription and WWW networks. 
Our findings that many directed networks are underlooped may have broad implications given that
such networks exhibit, for example, improved stability in foodweb systems \cite{foodweb} and
enhanced synchronization \cite{Syncrh} and transportation properties in various other systems \cite{Grad}.


\noindent{\it Short Loops in Random Networks}.
We first derive the expected number of 
self-avoiding loops in directed random networks. 
The general way to construct random uncorrelated undirected networks 
is by means of the Molloy-Reed model.  Given a set of nodes $V=\{i: 1,\dots,N\}$, 
the construction is based on 
generating a sequence of degrees $\{k^i\}$ from a given degree distribution
$P(k)$ with a structural cutoff $K={\cal O}(N^{1/2})$ \cite{cutoff}, and
randomly connecting the links.  In this ensemble, the expected number ${\cal
N}_L$ of short loops of length $L$ is given by \cite{loop_lungo,Burda} 
\begin{equation}
E_{\mbox{\small undir}}({\cal N}_L)=\frac{1}{2L}\left(\frac{\avg{k(k-1)}}{\avg{k}}\right)^L.
\label{random_und}
\end{equation}
This formula  implies that a network with diverging $\avg{k^2}$
has many  more short loops than networks with finite $\avg{k^2}$. In
particular,  scale-free networks with scaling exponent $\gamma\leq 3$ have many
short loops while Erd\H{o}s-R\'enyi networks have a negligible number of short
loops in the $N\rightarrow\infty$ limit. We now show that this expression can
be generalized to random {\it directed} networks. We again consider the
Molloy-Reed construction but in this case  we draw a sequence of incoming and
outgoing links $\{(k_{in}^i,k_{out}^i)\}$ from a degree distribution $P(k_{in},
k_{out})$  for all nodes $i\in V$. This distribution, which is not factorisable
in general, describes correlated variables $k_{in}$ and $k_{out}$ at any given
node. For directed uncorrelated networks, the structural cutoffs for in- and
out-degrees satisfy $K_{in}K_{out}<\avg{k_{in}}N$.
Proceeding as in the undirected case \cite{loop_lungo}, we obtain that the expected 
number of loops of size $L$ in the directed network ensemble is given
by 
\be
\label{random}
 E_{\mbox{\small dir}}({\cal N}_L)= \frac{1}{L}\left(\frac{\langle k_{in} k_{out}\rangle}{\avg{k_{in}}}\right)^L,
\end{equation}
where this approximate expression is valid for large $N$  and loop length satisfying $L\ll N {\avg{k_{in}k_{out}}^2}/{\avg{(k_{in}k_{out})^2}}.$
For undirected networks, $E_{\mbox{\small dir}}({\cal N}_L)$ reduces to $E_{\mbox{\small undir}}({\cal N}_L)$  because
the incoming connectivity is $k$ and the outgoing connectivity at the end point
of a link (on a self-avoiding loop) is $k-1$. The only difference is a factor 2, which accounts
for the orientation on the loops in Eq.~(\ref{random}). 

We observe from Eq.~(\ref{random}) that, in directed networks, the
one-point
correlation between the number of incoming and outgoing links modulates the
expected number of short loops. Indeed, if $k_{in}$ and $k_{out}$ on the same
nodes are not correlated, then the number of short loops is strongly reduced as
compared to the case when $k_{in}$ and $k_{out}$ are positively correlated.
The Barab\'asi-Albert (BA) networks \cite{ba}, for example, have small degree
correlations and are within the scope of
$E_{\mbox{\small dir}}({\cal N}_L)$ and  $E_{\mbox{\small undir}}({\cal N}_L)$ for uncorrelated random networks \cite{nota0}.
If we consider the undirected BA model, we find that the networks have
many short loops compared to random Erd\H{o}s-R\'enyi networks (in fact
$\avg{k(k-1)}\sim \log(N)$) \cite{Loops}. 
In contrast, if we consider the
directed version of the BA model (in which the incoming links are linked preferentially, and hence
$\avg{k_{in}k_{out}}=\avg{k_{in}}\avg{k_{out}}$), the networks have a
negligible number of short loops just as the Erd\H{o}s-R\'enyi networks in the
$N\rightarrow\infty$ limit.  

\noindent{\it Short Loops in a Given Network}.  A different approach is needed
for counting the loops of a {\it specific} directed network, as required in the
study of real systems.  In this case, as in the case of undirected networks
\cite{Loops}, the number of short loops can be expressed in terms of powers of
the adjacency matrix.  In particular, the number of (self-avoiding) loops of
length $L$ can be expressed as the total number of closed paths of length  $L$,
i.e. Tr $A^L/L$, minus the closed paths of length $L$ composed of
self-intersecting loops.  The number of loops of length $L$ in a network with
adjacency matrix $A$ is then given by  
$
{\cal N}_L=\frac{1}{L}\sum_{\{L_{\ell}\}}c(\{L_{\ell}\})\delta(L-\sum_{\ell} L_{\ell})
\sum_{i}\prod_{\ell}(A^{L_{\ell}})_{ii},
$
where the sequence $\{L_{\ell}\}$ describes the loop composition of the paths for
every correction term (for example, in the case $L=5$ we will find a correction
term involving paths composed of $\{L_{\ell}\}=\{2,3\}$ directed loops).  The
coefficients $c(\{L_{\ell}\})$  remain small for small $L$.

Starting from this general formula we derive upper and lower bounds for the
number of loops in a given directed network.  The upper bound is simply given
by the sum of all closed paths of length $L$, i.e.  ${\cal N}_L\leq\frac{1}{L}$ Tr
$A^L=\frac{1}{L} \sum_n \lambda_n^L$, where the sum is performed over all the
eigenvalues (including multiplicities). 
To find a lower bound we have to
express ${\cal N}_L$ in terms of the eigenvalues of the adjacency matrix
$A$ and in terms of its Jordan basis.  
In this way, it follows that ${\cal N}_L\simeq$ Tr $A^L/L$ 
provided that
$\kappa_L\equiv \max_{i}\sum_j \sum_m' \left(\begin{array}{c}L\\ m
  \end{array}\right)|\lambda_{j}^{-m}P_{ij}{P^{-1}}_{j+m, i}|
\ll 1,$
where $P$ is the matrix of generalized eigenvectors of $A$ in the Jordan
decomposition  $A=PJP^{-1}$ \cite{Syncrh} and  $\sum'_m$ 
indicates the sum over $m$ 
over the dimension of each Jordan block with associated eigenvalue  $\lambda_j$,
under the constraint that indices $j$ and $j+m$ are in the same block. 
If $\kappa_L\ll 1 $, the  dominant term in the expansion of ${\cal N}_L$ is the one with
$\{L_{\ell}\}=\{L\}$ and we have ${\cal N}_L\simeq\frac{1}{L}\sum_n\lambda_n^L$.

Comparing these results with the result found for the random case in
Eq.~\eqref{random}, it follows that a sufficient condition for a specific
network to have less short loops of length $L$ than its randomized versions is
$\sum_n\lambda_n^L < \left(\langle k_{in}
k_{out}\rangle/\avg{k_{in}}\right)^L$. Conversely, if $\kappa_L\ll 1$, a
condition for the network to have more loops is $\sum_n\lambda_n^L >
\left(\langle k_{in} k_{out}\rangle/\avg{k_{in}}\right)^L$. For loops in a
certain range of values $L\in(1, L_{c})$, it is convenient to restate these
conditions as 
\begin{equation} 
\overline{\lambda}\equiv \overline{(\sum_n
\lambda_n^L )^{1/L}} <\frac{\avg{k_{in}k_{out}}}{\avg{k_{in}}}
\label{suffunder}
\end{equation}
for the network to be under-shortlooped 
and
\begin{equation}     
\overline{\lambda}>\frac{\avg{k_{in}k_{out}}}{\avg{k_{in}}}\,\,\,\,
\mbox{if}\,\,\kappa=\max_{L\in(1,L_c)} \kappa_{L} \ll 1
\label{suffover}
\end{equation}
for the network to be over-shortlooped  {\it on average} over loop  lengths $L\in(1,L_c)$. 
The over-bar indicates average over  $L\in(1,L_c)$ for $L_c$
satisfying the condition for Eq. (\ref{random}) to be valid.

\noindent
{\it Long Loops}. The above analysis applies to short loops. 
Counting long loops is 
a difficult problem
for which approximate Monte Carlo \cite{Loopy} and statistical
mechanics methods \cite{BP} have been proposed in the undirected case.
To derive  a necessary condition for long {\it directed} loops to be present,
we use  percolation predictions \cite{Per_Boguna} for two-point correlated  
networks, where
the out-degree of a node is correlated (beyond the random condition)
with the in-degree of the nodes at the end points of its links.  
These networks are expected to account for the leading
correlation term that distinguishes a real network from its uncorrelated
random counterparts. 
In networks with two-point degree correlation, the percolation
condition for the largest strongly connected component (LSCC) is 
$\tilde{\Lambda}> 1$,
where $\tilde{\Lambda}$ is the largest eigenvalue of the two-point correlation
matrix $C_{\bf k', k}=(k_{out}P({\bf k'}|{\bf k}))$ \cite{Per_Boguna}. A strongly connected
component of a
network is a set of nodes where each node can reach and be reached by all the
others through directed paths.
For the uncorrelated random networks of the Molloy-Reed ensemble, the
largest eigenvalue of  matrix $C$ reduces to the known result 
$\tilde{\Lambda}=\frac{\avg{k_{in}k_{out}}}{\avg{k_{in}}}$. 
For a specific network, which is not necessarily well approximated by an
uncorrelated ensemble average, we can use the approximation
$\tilde{\Lambda}\simeq\Lambda$, where $\Lambda$ denotes the
largest eigenvalue of the adjacency matrix \cite{nota}. Consequently the
percolation conditions for the real and randomized networks are respectively $
\Lambda > 1 $ and $\frac{\avg{k_{in}k_{out}}}{\avg{k_{in}}}>1$.
Since the existence of a giant LSCC is a
necessary condition for the network to have long directed loops,
long loops are strongly suppressed when $\Lambda\leq 1$.
Because percolation only provides a necessary condition for the
existence of long loops, we make quantitative predictions
using a modified message-passing algorithm 
\cite{next} based on the belief propagation (BP) algorithm proposed in 
\cite{BP}. The algorithm provides an estimation for the  entropy $\sigma(L)=\log({\cal N}_L)/N$
of the loops of length $L$, from which we calculate ${\cal N}_L$. Within the conditions
discussed in \cite{next}, namely that the network is large and has a large number of loops,
this algorithm is able to predict the maximal loop length $L_{max}$ reliably.  However,
as shown below, the BP algorithm predicts correctly the under- or over-looped nature of all
networks in our database, including those with a small number of nodes or loops \cite{bp_vs_exact},
and the results are in very good agreement with the behavior suggested by the relative values
between $\Lambda$ and  $\frac{\avg{k_{in}k_{out}}}{\avg{k_{in}}}$.

\begin{figure}
\includegraphics[width=80mm, height=60mm]{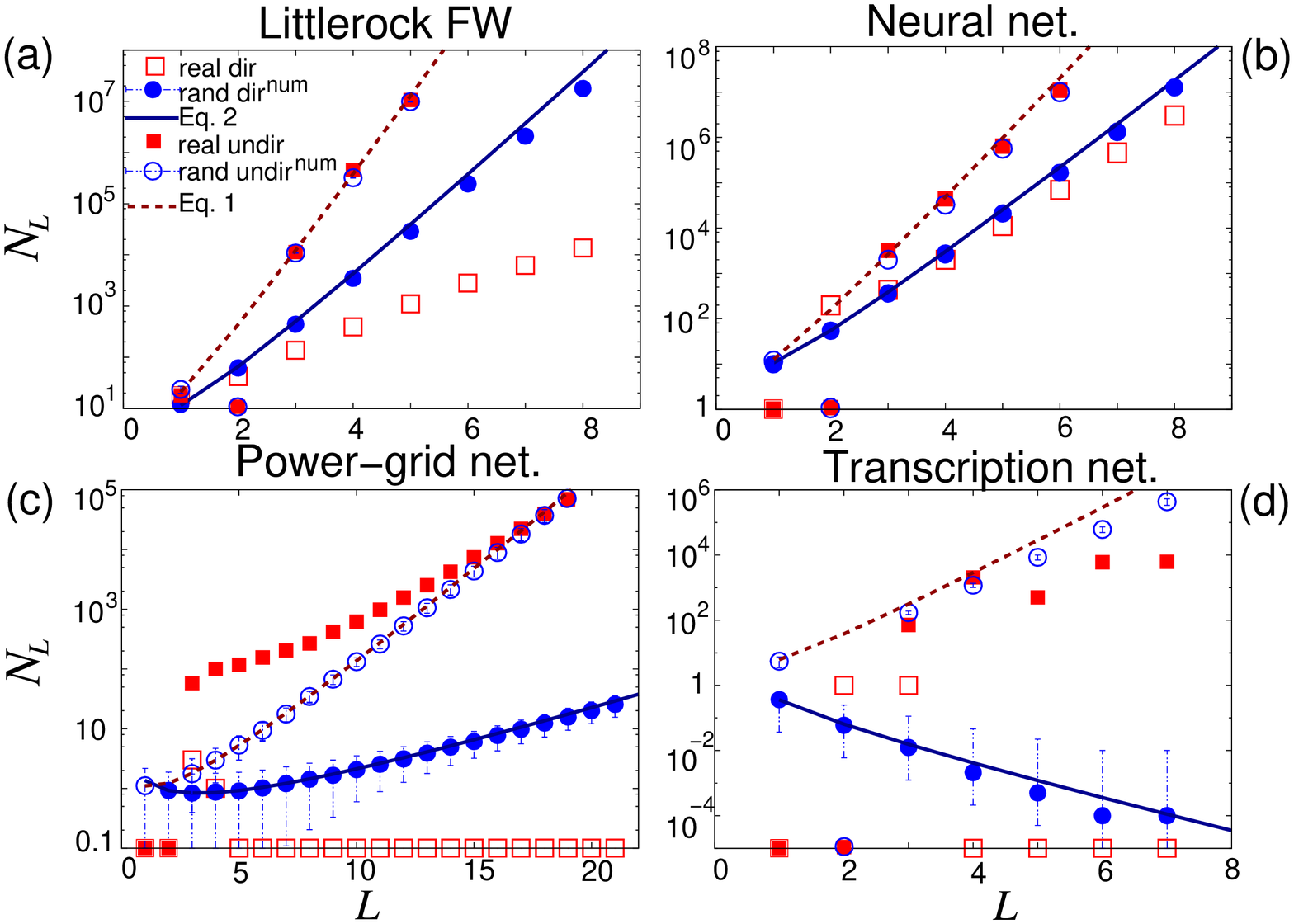}
\caption{\label{Loops.fig} 
(Color online) Number of short directed and undirected loops in several networks, where
different symbols correspond to the numerically determined values for the real and random counterparts of the networks. 
The lines indicate the theoretical predictions in Eqs.\ (\ref{random_und}) and (\ref{random}) for random networks.  
Points on the x-axis indicate no loops.} 
\end{figure}

\begin{table*}
\begin{ruledtabular}
\begin{tabular}{|l|c|c|c|c|c|c|c|c|}
 & \multicolumn{5}{c|}{Network Parameters} & \multicolumn{3}{c|}{Prediction/Actual$^{(a)}$}\\\hline
~~Network &N & M & $\Lambda$& $\frac{\avg{k_{in}k_{out}}}{\avg{k_{in}}}$  & $
\overline{\lambda}$  & Short loops$^{(b)}$ & Percolation/LSCC$^{(c)}$ & Long loops$^{(d)}$  \\\hline\hline
Littlerock FW & 183 & 2,494 & 7.00 &11.47 &7.93   &und/und 
&p-p/(12 vs 92) & und/und$^*$ \\\hline
Chesapeake FW&39 &177 &2.85&3.12  &2.40    &und/und 
&p-p/(41 vs 76) & und/und\\\hline
Mondego FW&46 &400 &8.95&9.14 &5.86       &und/und 
&p-p/(76 vs 92) & und/und$^*$ \\\hline
Seagrass FW&48&226&1.00&4.05   &1.65            &und/und 
&np-p/(0 vs 75) & und/und \\\hline
Metabolic net. & 532 & 596&2.85&2.58&3.00       &undet/over 
&p-p/(82 vs 94) & undet/undet$^*$ \\\hline
Power-grid net. & 4,889&5,855 &1.00&1.36&0.88           &und/und
 &np-p/(0.1 vs 33)& und/und\\\hline
ND WWW      &325,729& 1,497,135&152.00 &43.14&153.32     & over/over
 &p-p/(17 vs 41) & ---\\\hline
Neural net. &306& 2,359&9.15&10.49 & 8.84       & und/und 
&p-p/(78 vs 86) &  und/und$^*$ \\\hline
Transcription net. & 688& 1,079&1.32&0.36 & 0.88       & undet/over 
&p-np/(0.4 vs 0.3) & over/over 
\end{tabular}
\end{ruledtabular}
\caption{Properties of real directed networks: number of nodes $N$ and links $M$, eigenvalue $\Lambda$,
$\avg{k_{in}k_{out}}/\avg{{k_{in}}}$, and spectral quantity  $\overline{\lambda}$  (l.h.s.\ columns);
loop structure and percolation properties (r.h.s.\ columns). The values of
$\kappa$ for the undetermined and over-shortlooped cases are 96.0, 73.2 and 0.2 for the
metabolic, transcription and WWW network, respectively.
$^{(a)}$Underlooped (und), overlooped (over), undetermined (undet), not determined numerically (---).
$^{(b)}$Actual values determined by averaging over the directed loops up to length $L_c = 6$ 
($L_c= 3$ for the ND WWW). 
$^{(c)}$From left to right: predicted percolating (p) or non-percolating (np) LSCC in real and random networks together with the
actual percentage of nodes in the LSCC of the real vs.\ random networks.
$^{(d)}$From left to right: prediction for long loops obtained using
the BP algorithm \cite{next} to estimate $L_{max}$ and the actual result
obtained using exhaustive or partial ($^{*}$) enumeration of the loops.
\label{table}}
\end{table*}

\begin{figure}
\includegraphics[width=80mm, height=60mm]{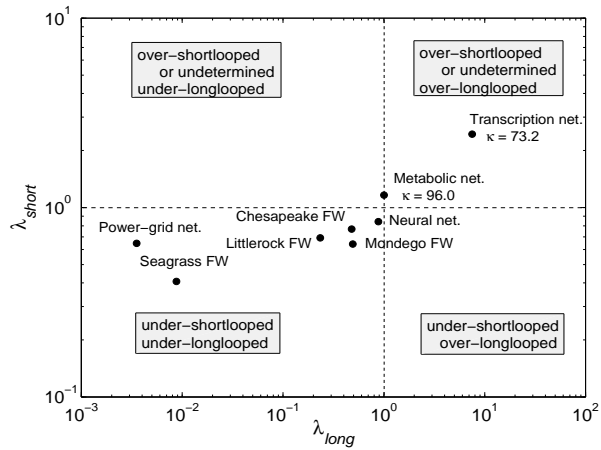}
\caption{Underlooped, overlooped and undetermined regions
in the $\lambda_{short}\equiv \overline{\lambda}/(\avg{k_{in}k_{out}}/\avg{{k_{in}}})$ 
vs.\ $ \lambda_{long}\equiv L_{max}^{real}/\avg{L_{max}^{rand} }$ diagram, where $L_{max}^{real}$
and $L_{max}^{rand}$ are predicted using the BP algorithm.  The points correspond
to the predictions for both short and long loops for the networks in Table I, except for the
ND WWW, which is over-shortlooped and is not shown because it is difficult to calculate its
$ \lambda_{long}$. The actual counting of the loops confirms
the predictions (Table I).
\label{diag}}
\end{figure}

\noindent{\it Real Networks}. We consider several real directed networks
\cite{nnets}: ({\it i}) Texas power grid; ({\it ii}) foodwebs (Chesapeake, Mondego,
Littlerock, and Seagrass regions); ({\it iii}) metabolic network of 
{\it E. coli}, where the nodes represent metabolites; ({\it iv}) 
Notre Dame University's WWW; ({\it v}) {\it C. elegans}' 
neural network; and ({\it vi}) transcription network of {\it S. cerevisiae},
where the nodes correspond to regulating and regulated genes. 
Figure~\ref{Loops.fig} shows the distributions of short loops (measured using exact
enumeration \cite{Tarjan}) for both the directed and undirected versions  
of four real networks along with the randomized counterparts of same number of
in- and out-links in each node.
The randomized networks 
are well approximated by the theoretical predictions in Eqs. (\ref{random_und}) and (\ref{random}), 
as indicated by the lines in the figure.
Directed networks tend to have less loops
than undirected networks, as expected. However, while real undirected
networks tend to have more loops than random ones, the opposite occurs in 
the directed case. 

Indeed, six out of the nine directed
networks we analyzed are under-shortlooped,
as shown in Fig.~\ref{diag} and Table I.
The only exceptions are the metabolic
and transcription networks, which are marginally over-shortlooped, and
the WWW network, which is the only social network present in our database \cite{com_www}.
These findings are very different from what one would anticipate from previous
studies on undirected networks, where highly clustered small-world networks
prevail. Figure~\ref{diag} summarizes the 
prediction and actual tendency of the directed networks to be underlooped.
Table I summarizes the network parameters and results for all  directed networks
analyzed, where  $\overline{\lambda}$ is calculated by summing over all loops up to a
length cutoff $L_c$ chosen to be $6$ \cite{com_www}. 
Our predictions compare well with direct data analysis.

\noindent{\it Conclusions}.
We have studied deviations in the loop statistics 
and provided criteria for
determining if a network is underlooped or overlooped
compared to its randomized
counterparts. Empirical evidence coming from the study of different types of natural and
technological networks shows that many of these different networks are under-shortlooped,
a surprising result which is in sharp contrast with the tendency of undirected networks
to be over-shortlooped. The only socio-technological network in our databse, the ND WWW,
contains instead very many short loops. We expect that our results will be important and
further extended in the study of social, biological and technological systems.
In social networks, the abundance of directed loops can be an important factor in the promotion
of mutual reinforcement amongst agents \cite{social}, while in cellular and neural networks it
can play a major role in information processing \cite{neuron} and regulation \cite{cell}. 
In other systems, the reduced number of directed loops can lead to improved
stability \cite{Syncrh,foodweb} and transportation properties \cite{Grad}, which we hope will
stimulate other applications of our findings.


The authors thank Dong-Hee Kim and Marian Bogun\'a  for providing feedback on the manuscript.
This work was supported by IST STREP GENNETEC Contract No.\ 034952 (GB),
DOE LANL Contract No.\ DE-AC52-06NA25396 (NG), and NSF award DMS-0709212 (AEM).


\begin{thebibliography}{99}
\bibitem{corr}
We consider degree correlations \cite{Alex,Per_Boguna,Simp_paradox} 
of two types: between the in-out degrees of a given node and between
the degrees of connected nodes.
\bibitem{nota}
Because Tr$\, C^L\simeq$ Tr$\, A^L$, if the two matrices are expanders
we have $\tilde{\Lambda}\simeq \Lambda$.
\bibitem{bp_vs_exact}
For example, $L_{max}^{real}/\langle L_{max}^{rand}\rangle = 0.48$ (BP) and $0.19$ (exact)
for the Chesapeake network while $L_{max}^{real}/\langle L_{max}^{rand}\rangle<0.004$ (BP) and $<0.08$ (exact)
for the power grid, consistently indicating that long loops are underrepresented in these networks.
\bibitem{com_www}
For the ND WWW, due to its size, we use $L_c=3$ and we estimate 
$\overline{\lambda}$ and $\kappa$ from the 100 largest eigenvalues
and corresponding eigenvectors.
\end{thebibliography}
\end{document}